# The Effects of Transverse Inclination on Aeroelastic Cantilever Prisms---Phenomenology, Unsteady Force, and the Base Intensification Phenomenon

Zengshun Chen [a], Jie Bai [a], Cruz Y Li [b,*], Yemeng Xu [a], Jianmin Hua [a], Xuanyi Xue [a,*]

a School of Civil Engineering, Chongqing University, Chongqing 400045, China

b Department of Civil and Environmental Engineering, The Hong Kong University of Science and Technology, Clear Water Bay, Kowloon, Hong Kong, China.

Corresponding author: Cruz Y Li, Xuanyi Xue

mail address: yliht@connect.ust.hk

Mailing address: School of Civil Engineering, Chongqing University, Chongqing 400045, China

**Abstract**

The transverse inclination is a probable scenario when inclined structures experience an inflow of altered attack angles. This work investigates the effects of transverse inclination on an aeroelastic prism through forced-vibration wind tunnel experiments. The aerodynamic characteristics are tri-parametrically evaluated under different wind speeds, inclination angles, and oscillation amplitudes. Results show that transverse inclination fundamentally changes the wake phenomenology by impinging the fix-end horseshoe vortex and breaking the separation symmetry. The aftermath is a bi-polar, once-for-all change in the aerodynamics near the prism base. The suppression of the horseshoe vortex unleashes the Kármán vortex, which significantly increases the unsteady crosswind force. After the initial morphology switch, the aerodynamics become independent of inclination angle and oscillation amplitude and depend solely on wind speed. The structure's upper portion does not feel the effect, so this phenomenon is called *Base Intensification*. The phenomenon only projects notable impacts on the low-speed and VIV regime and is indifferent in the high-speed, quasi-steady Galloping regime. In practice, Base Intensification will disrupt the pedestrian-level wind environment from the unleashed Bérnard-Kármán vortex shedding, making it erratic and gusty. Moreover, it increases the aerodynamic load at a structure base by as much as 4.3 times. Since fix-end stiffness prevents elastic dissipation, the load translates to massive stress, making detection trickier and failures, if they are to occur, more sudden, extreme, and without any warnings. The 4.3-time amplification also surpasses the safety factor of many standard designs, so transverse inclination deserves engineering attention.

**Keywords**: Forced-Vibration Wind Tunnel Experiment; Transverse Inclination; Unsteady Aerodynamic Force; Aerodynamic Damping; Horseshoe Vortex; Vortex Shedding.

**Highlights**

- ✓ A series of forced-vibration tests performed to obtain unsteady aerodynamic forces acting on transversely inclined slender prisms;

- ✓ Unsteady aerodynamics characteristics on transversely inclined slender prisms analyzed;

- ✓ The aerodynamic damping of the inclined prism identified;

- ✓ Suppression of horseshoe vortex and the base intensification phenomenon studied.

# 1. Introduction

With the continuous improvement of construction technology, modern buildings are becoming increasingly slender and flexible. Consequently, the influence of wind loading and wind-induced vibration on super-tall structures is of great safety concern, particularly after being reminded of the recent vibration of the SEG Tower in Shenzhen. Therefore, studies on the aerodynamic characteristics of super-tall buildings bear critical importance. To date, much remain unexplored for even the most canonical configurations, for example, the square prism.

The aerodynamic characteristic of a structure is the key determinant of its wind-induced response. In the event of excessive excitations, whether by the Vortex-Induced-Vibration (VIV), galloping, or fluttering, irreversible damage may be incurred on the structure(Blevins and Saunders, 1979; Païdoussis et al., 2013). Therefore, a class of numerical(Li et al., 2020; Zhang et al., 2022, 2020), experimental(Hu et al., 2017; Mannini et al., 2015a, 2017), and even data-driven(He et al., 2022; Li et al., 2022; Raissi et al., 2019) investigations has been performed on bluff-body aerodynamics. For example, (Tanaka et al., 2012) studied the aerodynamic characteristics of unconventional building models (spiral, cone, *etc.*) through rigid pressure test. (Hui et al., 2019) examined the influence of different façade appendages on the aerodynamic characteristics of high-rise buildings. Likewise, (Carassale et al., 2014) studied the influence of round corners on the aerodynamic characteristics of rectangular prisms.(Li et al., 2021a) formulated a data-driven architecture to deterministic associate fluid excitation and structural pressure.

However, these studies are limited to the scenarios that the structure is aligned vertically perpendicular to the incoming flow, which might not be the case in modern designs. To this end, (Hu, et al., 2015) studied the longitudinally inclined (forward and backward) rectangular prism through a series of wind tunnel tests, concluding the profound and vastly different effects of inclination on the prism's aerodynamic characteristics depending on the inclination angle. Perhaps the greatest limitation of this work is the rigidity of the test model. Due to the complexities of fluid and its interactive mechanisms with a structure, evaluating the unsteady aerodynamic force with accuracy is immensely difficult through a rigid test. The inclusion of aeroelasticity is necessary to accommodate fluid-structure interaction.

Supporting this notion, (Hu et al., 2015) later confirmed the quasi-steady theory inadequate for predicting inclined structural responses using an aeroelastic model.(Mannini et al., 2015b), too, disproved the theory for low mass-damping structures. On this note, (Chen et al., 2021a) also observed the failure of quasi-steadiness in predicting a tapered prism's motion, through which they discovered the *Partial Reattachment* phenomenon. Research showed that structures with shape irregularities generally disobey the quasi-steady theory. In other words, ignoring the unsteady effects brought about by aeroelasticity is equivalent to a quasi-steady prediction, which may work for some regular geometries but is bound to deviate from reality with irregular ones.

To this end, previous research proved that the forced-vibration method produces relatively accurate evaluations of the unsteady effects (Chen et al., 2018b; Cooper et al., 1997; Zou et al., 2020). (Kim et al., 2018) examined the impact of vibration amplitude on the aerodynamic forcing, finding notable improvements of forced-vibration compared to the rigid test. Furthermore, (Chen et al., 2020, 2021c) systematically compared the aerodynamic forces measured from rigid, forced-vibration and the hybrid aeroelastic-pressure balance (HAPB) test (*i.e.,* free-vibration). In addition, they (Chen et al., 2017) found that the overwhelming effects of vibration act only on the crosswind faces. The proportionality between lift and vibration also only exists at low wind speeds. The inclination angle may also lead to significant deviations from the vertical case. On this note, vibration amplitude, wind speed, and inclination angle are three key factors of investigation.

The thorough literature review discloses an alarming gap in bluff-body aerodynamics: there is yet a study on the transversely inclined prism with aeroelasticity. However, the configuration is highly probable in real life, given a change in wind attack angle on any inclined structure. Perhaps the only referential effort was the one that proposed a modified quasi-steady model for transversely inclined prisms (Chen et al., 2018a). However, the predicted response was only acceptable for the onset of galloping but notably off thereafter. Therefore, the aerodynamic characteristics of transverse inclination and the feasibility of the quasi-steady theory demand immediate attention.

This work aims to examine the aerodynamic characteristics of transversely inclined prisms. The unsteady aerodynamic forces acting on the model were obtained through a series of forced-vibration tests, and the aerodynamic characteristics of the model were analyzed under representative wind speeds, inclination angles, and vibration amplitude. In composition, Section 1 offers contextual information and a thorough literature review. Section 2 details the methodology of the forced-vibration wind tunnel test. Section 3 presents the model's aerodynamic characteristics from test results and elucidates the influence of inclination angle. Section 4 discusses the aerodynamic damping of transversely inclined slender prisms. Section 5 summarizes the major findings of this work.

## 2. Wind tunnel tests and tests models

The wind tunnel tests were performed in the high-speed section of the CLP Power Wind/Wave Tunnel Facility at the Hong Kong University of Science and Technology. The dimensions of the wind tunnel are 29.2 m (length) × 3 m (width) × 2 m (height). The corresponding blockage ratio was 0.78%, with which undesired influence of tunnel walls on the wind field was avoided (Holmes, 2018). The mean wind profile exponent $\gamma$ was 0.15, and the turbulence intensity at the model top was 10%. The corresponding provisions were compared with the measured wind characteristics, as shown in Fig. 1 (a) and (b). The comparison shows that the simulated wind characteristics closely replicate the target ones, therefore, appropriate for wind tunnel testing.

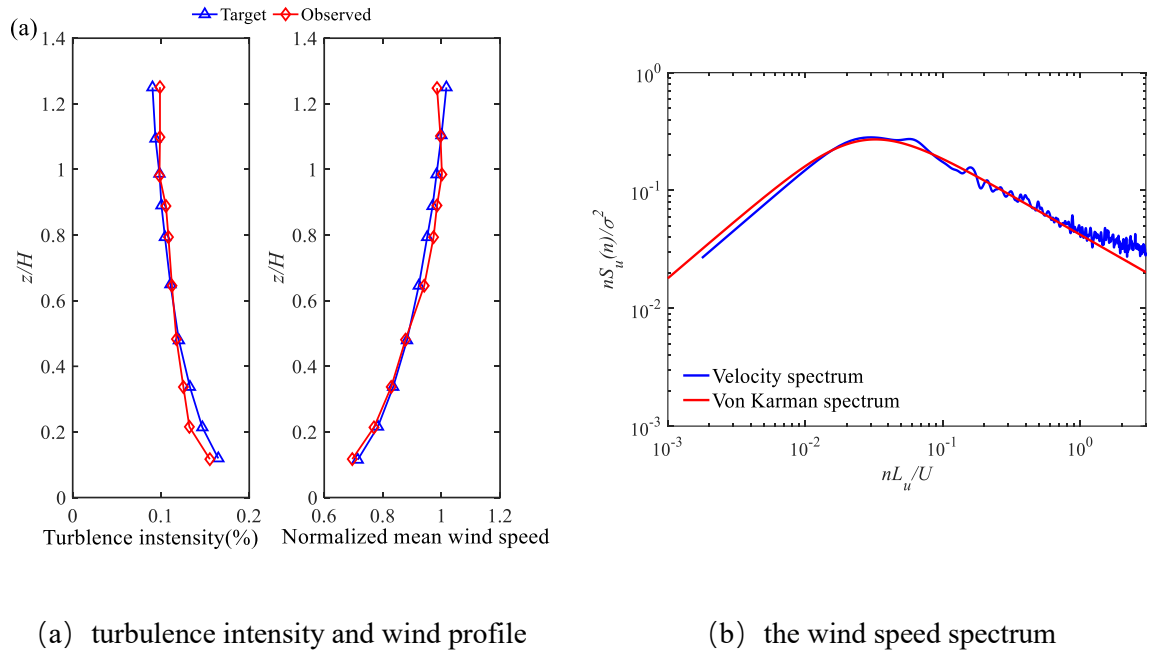

(a) turbulence intensity and wind profile  (b) the wind speed spectrum

**Fig. 1** Wind field characteristics.

The test model was a square prism with the dimensions 50.8 mm(*B*)×50.8 mm(*D*)×915 mm(*H*), where *B* and *D* are the cross-sectional length and width, and *H* is the height (see Fig. 2). The aspect ratio of the model is 18:1. The vibration frequency of the model was set to 7.8 *Hz*. In the forced-vibration test, a vibrating device under the model base instigated harmonic excitations. The inclination angle *α* of the model varied from 0° to 30°, and the amplitude of oscillation varied between 6% and 20%.

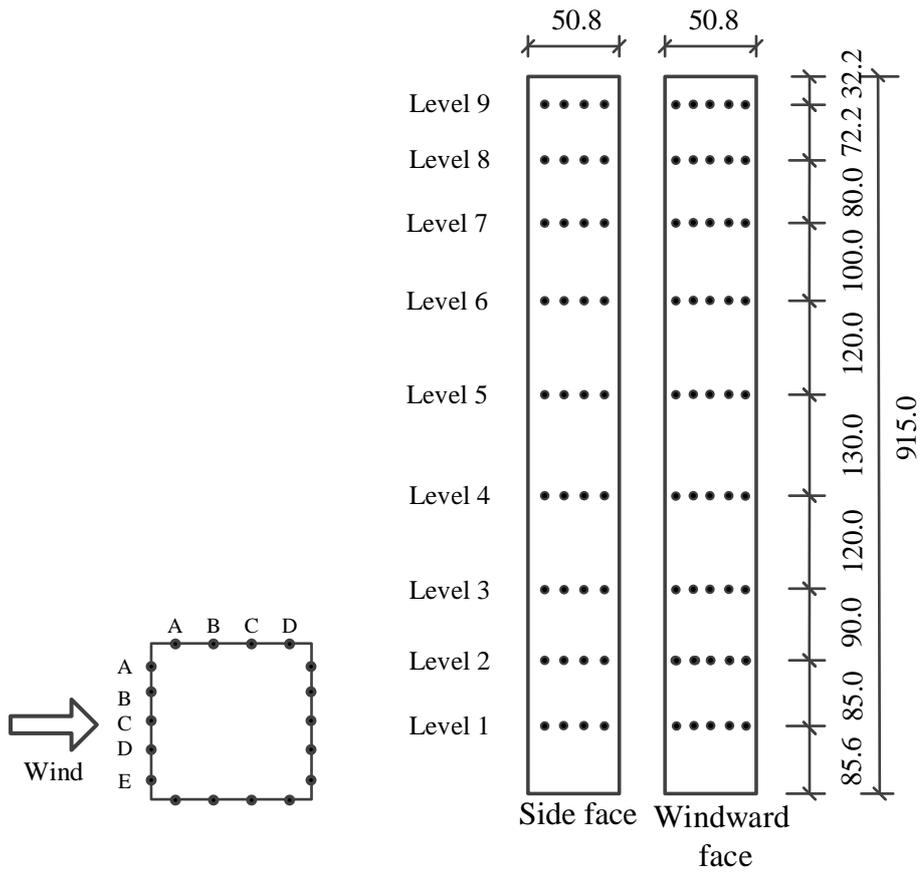

(a)

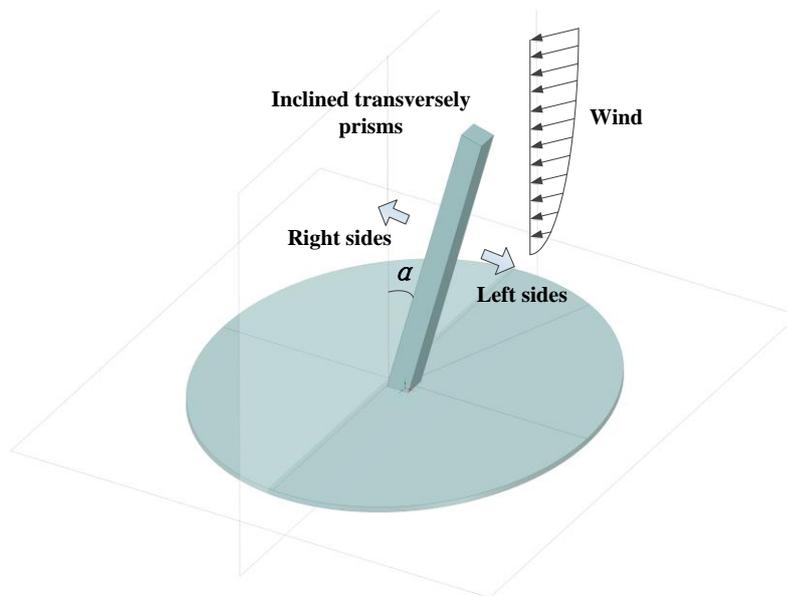

(b)

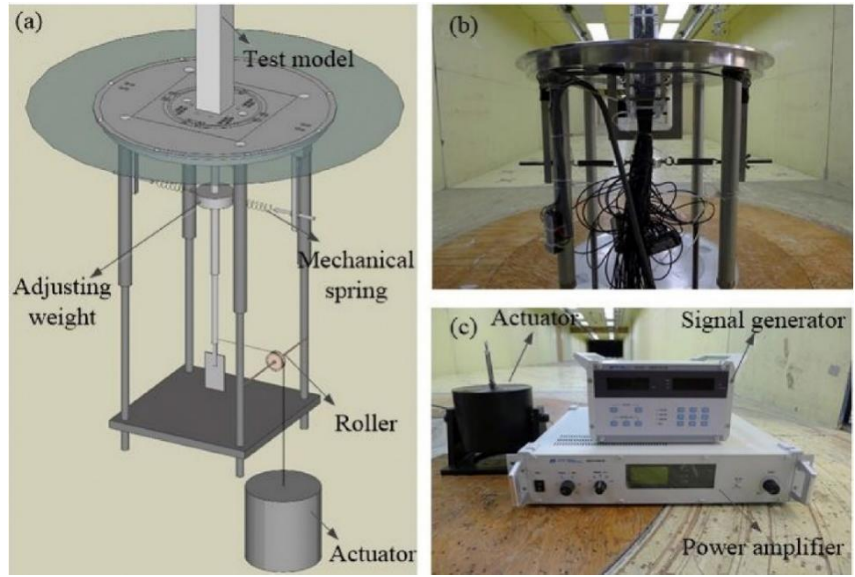

(c)

**Fig. 2** Model for pressure measurements: (a) The distribution of Pressure taps; (b) transversely inclined prisms; (c) The forced-vibration device and system

The reduced wind speed $V_r = U/fD_c$ ($U$ is the wind speed at the top of the model, $f$ is the natural frequency, and $D_c$ is the characteristic length) was set between 6 and 20. More experimental details are summarized in Table 1. The sampling frequency and the duration were set to 500 $Hz$ and 100$s$. The tip response of the model was obtained by correcting the raw data measured from the strain gauge at the model base. A Multi-Point Synchronous Pressure Measurement System (SMPSS) measured the unsteady wind pressures, in which the distribution of the Pressure Taps is illustrated by Fig. 2(a). The pressures and response measurements are also in synchronization. To avoid literary redundancy, readers may refer to more details in (Chen et al., 2018b), which adopted the same testing parameters.

Table 1 Basic parameters of the model and wind tunnel test

| Height (mm) | Number Nodes | Reduced Speeds | Vibration Amplitude (%) | Inclination Angle $\alpha$ (°) | $\gamma$ | Cross-Section (mm) |
|---|---|---|---|---|---|---|
| 915 | 162 | 6-8, 13-20 by increment of 1<br>8-12 by increment of 0.5 | 0, 8-20 by increment of 2 | 0-30 by increment of 5 | 0.15 | 50.8×50.8 |

## 3. Results and Discussion

### 3.1 Validation of the pressure measurement

To validate the unsteady pressure measurement, the forced vibration test ($\alpha = 0°, V_r = 18$, $\sigma_y/D = 0\%$) is compared to a previous study (Chen et al., 2017). The altitudinal RMS lift coefficient $C_{L,\text{rms}}(z)$, altitudinal RMS drag coefficient $C_{D,\text{rms}}(z)$ and altitudinal mean drag coefficient $C_{D,\text{mean}}(z)$ are expressed as:

$$C_{L,\text{rms}}(z) = \frac{2\tilde{F}_L(z)}{\rho A(z) U^2} \quad (1)$$

$$C_{D,\text{rms}}(z) = \frac{2\tilde{F}_D(z)}{\rho A(z) U^2} \tag{2}$$

$$C_{D,\text{mean}}(z) = \frac{2\bar{F}_D(z)}{\rho A(z) U^2} \tag{3}$$

where $\tilde{F}_L(z)$ is the local RMS lift force, $\tilde{F}_D(z)$ and $\bar{F}_D(z)$ are the local RMS drag force and the local mean drag force; $A(z)$ is the local area at height $z$. In **Error! Reference source not found.**, the local mean and RMS force coefficients observed in the present study are in close agreement with a previous study. The results in **Error! Reference source not found.** indicate that the pressure measurement is reliable and can be utilized for further analysis.

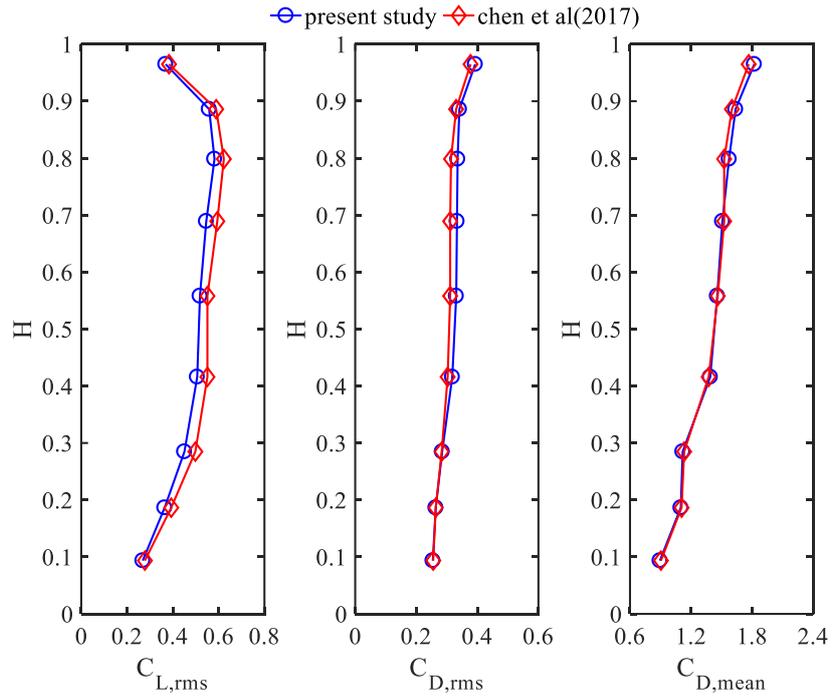

**Fig.3.** Comparisons of between the altitudinal force coefficients obtained in the present study and those in literature.

## 3.2 Pressure, Force, and Phenomenology

*3.2.1 Mean Pressure Distribution*

Equation (4) was used to calculate the mean wind pressure coefficient $C_P$ acting on the model:

$$C_P = \frac{2P_i}{\rho U^2} \qquad (4)$$

where $P_i$ is the altitudinal pressure acting on the model and $\rho$ is the density of air.

Due to the change of wind attack angle, the flow separation at the leading edge is not symmetric. Therefore, the most representative observations come from the side faces. Given the abundance of test results, the most insightful sets of data, with wind speeds of $V_r = 11$ and $V_r = 18$ and vibration amplitudes $\sigma_y/D$ = 0%, 8%, and 18%, are presented. $V_r = 11$ is characteristic of the lock-in region and $V_r = 18$ is characteristic of the high-speed galloping region. As shown in Fig. 2 (b), in subsequent discussions, the left side refers to the near-wind side that faces the ground. The right side refers to the far-wind side that faces the ceiling.

We begin by examining the side faces. Fig.4 shows the pressure distribution at $\alpha = 0°$, which, regardless of lock-in or galloping, is nearly symmetric with only minor disparities. The observation meets anticipation and verifies the test's accuracy. Both faces display mild and symmetric signs of reattachment near the base's rear edges.

By sharp contrast, Fig.5 demonstrates the marked effect of transverse inclination in the lock-in regime. On the left face, intense low suction zones form and propagate towards upstream and higher altitudes from the rear corner. This is a clear indicator of intensified reattachment.

However, very interestingly and contrary to intuition, the intensification at small $\alpha$ is stronger than larger $\alpha$ and is uniformly thwarted near the mid-span. There seems to be a powerful initial intensification but consistency with $\alpha$ afterward. Since reattachment signals the impingement of shear layers' momentum thickness and the reverse flow's closure into separation bubbles, the observations trace back to the shear layer and vortical activities.

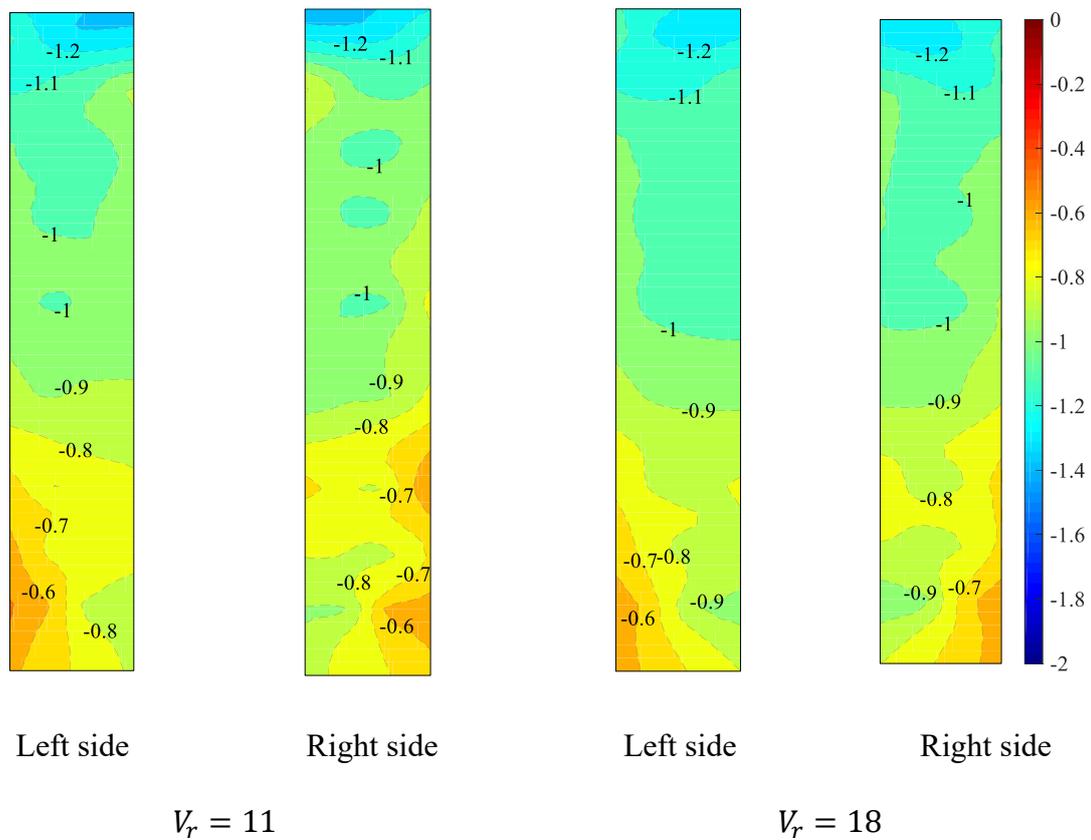

| Left side | Right side | Left side | Right side |

$V_r = 11$  $V_r = 18$

**Fig.4** Mean pressure distribution on the side face ($\alpha = 0°$ $\sigma_y/D = 0\%$).

The clear difference between the left and right faces attests to the broken symmetry after inclination. The intensification on the right face is much milder. Nevertheless, the variation patterns are consistent. After the initial intensification from $\alpha = 0°$ to $10°$, the base pressure remains unchanged with a further increase of $\alpha$. The prism's upper half also barely sees the effect of inclination.

By bi-parametrically varying $\alpha$ and $V_r$, Fig.6 presents pressure distribution in the galloping regime. From VIV to galloping, the effect of wind speeds or mode of vibration is anything but cataclysmic. The suction zones are generally more concentrated and restricted to the model base. The promoted locality of reattachment also means nearly 2/3 of the upper prism is unaffected by inclination. Pressure extremities remain unchanged apart from a slight intensity increment on the right face (minima from -0.5 to -0.4).

*3.2.2    The Base Intensification Phenomenon*

The similitude between the VIV and galloping regimes reveals an important conclusion. The fundamental difference between VIV and galloping is their fluid mechanics, or more specifically, flow phenomenology. The former arises from the resonance between the Bérnard-Kármán vortex shedding with a structure's natural mode of vibration(Flemming and Williamson, 2005). The latter arises from inflow instabilities and geometric asymmetry, which has little to do with the resonance of the Kármán Street(Hu et al., 2016; Parkinson, 1989). So, the intensification's indifference to the regime transition means it is not phenomenologically related to the Kármán vortex(Unal and Rockwell, 1988a, 1988b).

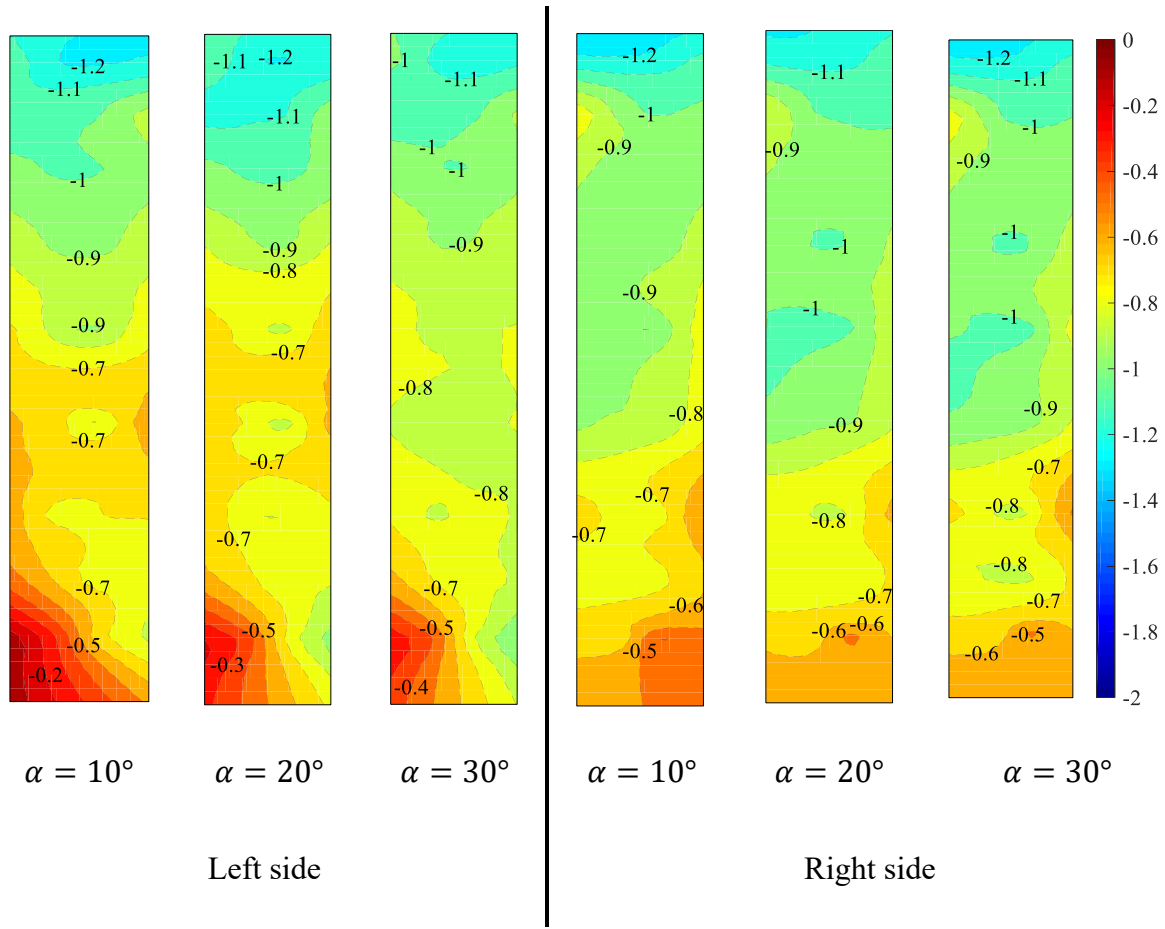

**Fig.5** Mean pressure distribution on the side face with different inclination angles

($V_r = 11$).

Moreover, Kármán shedding takes place at all altitudes except being suppressed near the two ends, but the base is precisely where the intensification occurs, not to mention transverse inclination affects only the prism's bottom half. So, the observed reattachment is not the typical type, which impinges the leading-edge vortex, one observes from vertical structures. Instead, this phenomenon impinges the fix-end horseshoe vortex. To this end, a concept central to this paper is clarified: transverse inclination confines the available space between the near-wind wall and the ground, squeezes passage of flow, and impinges the horseshoe vortex. We refer to this asymmetric morphology as the *Base Intensification* phenomenon.

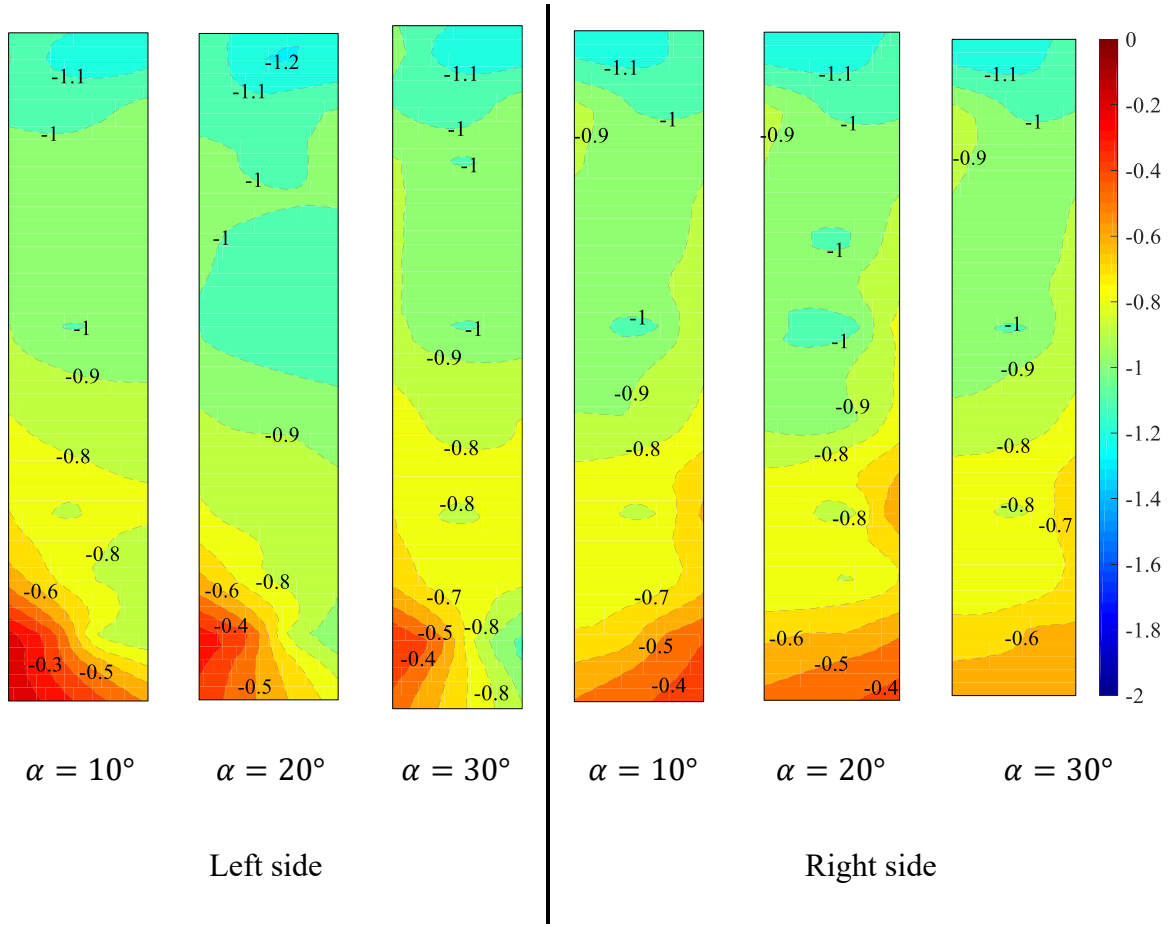

**Fig.6** Mean pressure distribution on the side face with different inclination angles ($V_r = 18$).

Previous discussions demonstrated that the Base Intensification is, once incurred by the initial inclination from $\alpha = 0°$ to $10°$, indifferent to inclination angle and the VIV-Galloping regime change. Fig.7 and Fig.8 further elucidate its relationship with vibration amplitude. In both the VIV ($V_r = 11$) and Galloping ($V_r = 18$) regimes, and at the most intense angle $\alpha = 10°$, varying the vibration amplitude $\sigma_y/D$ from 0-18% only minimally affects the pressure distributions.

Therefore, the Base Intensification phenomenon is also independent of vibration amplitude.

These empirical observations are fascinating because the Base Intensification seems to be a once-and-for-all switch of flow morphology after initially prescribing a transverse inclination. The explanation is that symmetry is a bi-polar state---a configuration can only be symmetric or asymmetric; there is no intermediate 'trans-symmetric' state. Therefore, the initial intensification is the direct result of asymmetric separation. As for why the Base Intensification is indifferent to other factors, our conjecture is that wind speed, inclination angle, and vibration amplitude are all determinants of shear layer dynamics. However, unlike the Kármán vortex, the horseshoe vortex is not a result of shear layer activities. Its sensitivity to inflow parameters is, therefore, frigid.

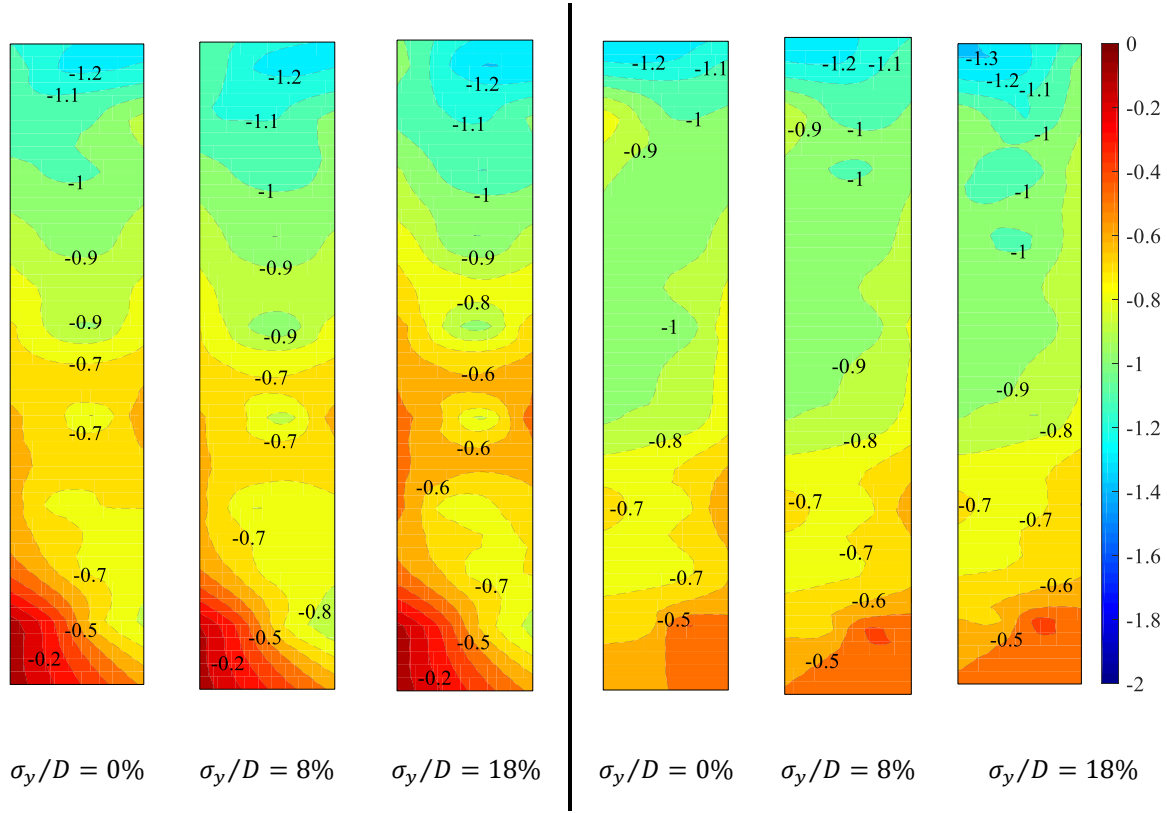

$\sigma_y/D = 0\%$  $\sigma_y/D = 8\%$  $\sigma_y/D = 18\%$ | $\sigma_y/D = 0\%$  $\sigma_y/D = 8\%$  $\sigma_y/D = 18\%$

Left face | Right face

**Fig.7** Mean pressure distribution on both sides with different amplitude($V_r = 11$  $\alpha = 10°$).

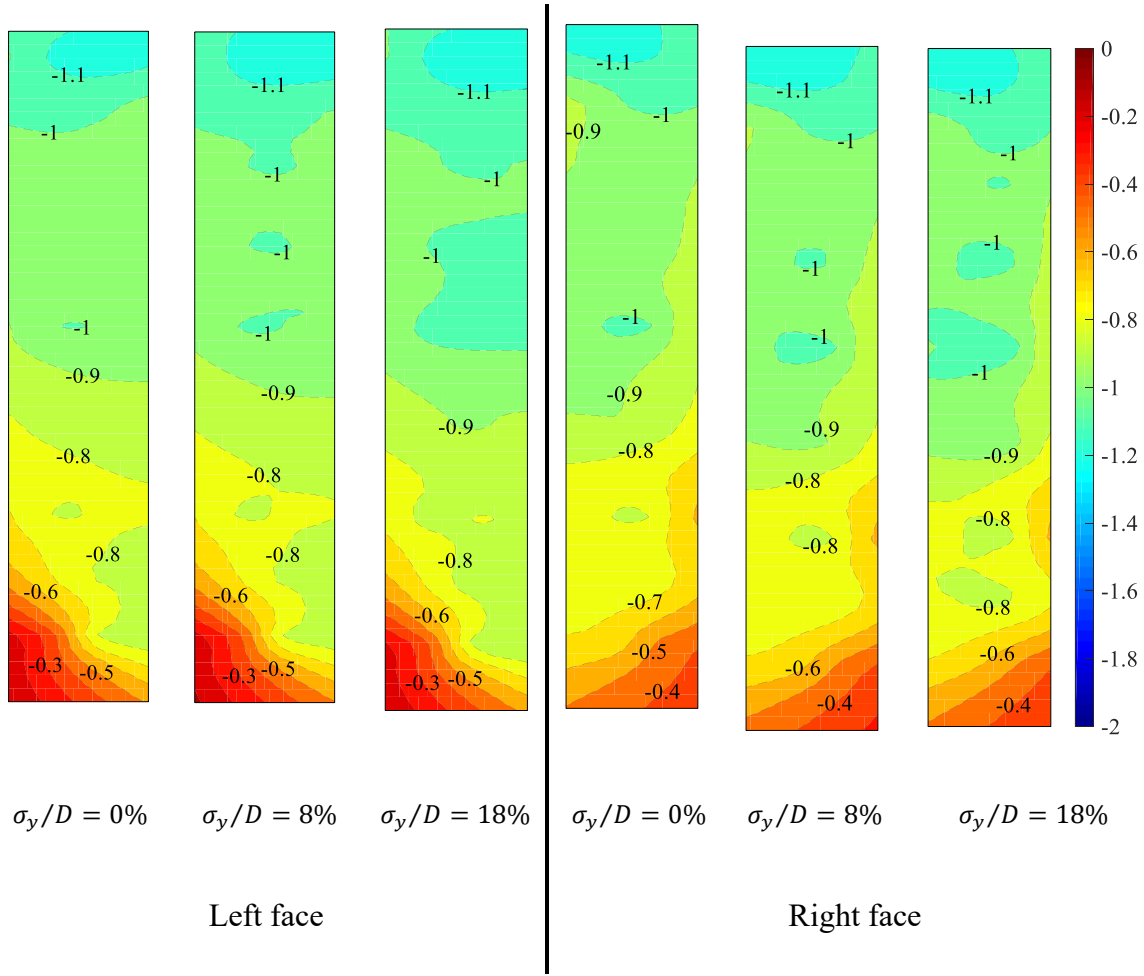

$\sigma_y/D = 0\%$   $\sigma_y/D = 8\%$   $\sigma_y/D = 18\%$   $\sigma_y/D = 0\%$   $\sigma_y/D = 8\%$   $\sigma_y/D = 18\%$

Left face | Right face

**Fig.8** Mean pressure distribution on both sides with different amplitude($V_r = 18$ $\alpha = 10°$).

*3.2.3  Generalized Lift*

Following the phenomenological analysis, we proceed to evaluate how Base Intensification affects the aerodynamic parameters. The unsteady aerodynamic force coefficient in the crosswind direction is defined as:

$$\tilde{C}_{L,rms} = \frac{\int_0^H A(z) C_{L,rms}(z) \phi(z) dz}{DH} \tag{5}$$

where $\tilde{C}_{L,rms}(z)$ is altitudinal RMS lift force coefficient. Equation (5) is obtained by integrating Equation (1) along the model height, in which $A(z)$ is the altitudinal area at height z, $\phi(z)$ is the function of the first mode, namely $\phi(z) = z/H$.

Since the mean lift coefficient is approximately zero, only the RMS lift bears implications (Lin et al., 2005). Fig.9 presents the tri-parametric variation of the globally generalized RMS lift coefficient at different α, Vr, and σy/D. We first examine the effect of α. Clearly, the variation pattern is largely a concave parabola in the VIV regime at α = 0°. But those at α = 5°-30° take shape of an exponential decay. The bi-polar nature of Base Intensification is once again reflected in the lift coefficient.

Due to transverse inclination, the maximum lift for all σy/D increases. The largest increase is ~0.2 at σy/D = 0%, which proportionately reduces to ~0.1 at σy/D increases to 20%. The observation is telling in two aspects. The observation is telling in two aspects. First, Base Intensification significantly promotes VIV by increasing its maximum lift and shifting its maxima to a lower Vr, rendering it perilous for structural safety. Second, the greatest lift increase occurs with no vibration amplitude, making the activation of VIV easier from the state of rest. The Base Intensification phenomenon is a potentially severe detriment.

Moreover, the inclined cases display a clear self-similarity in the VIV regime, where curves translate up with increasing σy/D. This translation appears rather simplistically as a linear superposition to reflect additional kinetic energy injected into the system by enhanced forced-vibration. This linearity in the highly nonlinear regime of the Kármán vortex dissociates the Base Intensification from the shear layer activities, and reinforces its indifference to vibration amplitude. On the other hand, the convergence of the $\alpha = 0°$ curves in the Galloping regime meets anticipation because this is where the quasi-steady theory thrives. However, the convergence is also observed for all inclined cases and even with the same lift ~0.25. To this end, the Base Intensification affects only the VIV but not the Galloping.

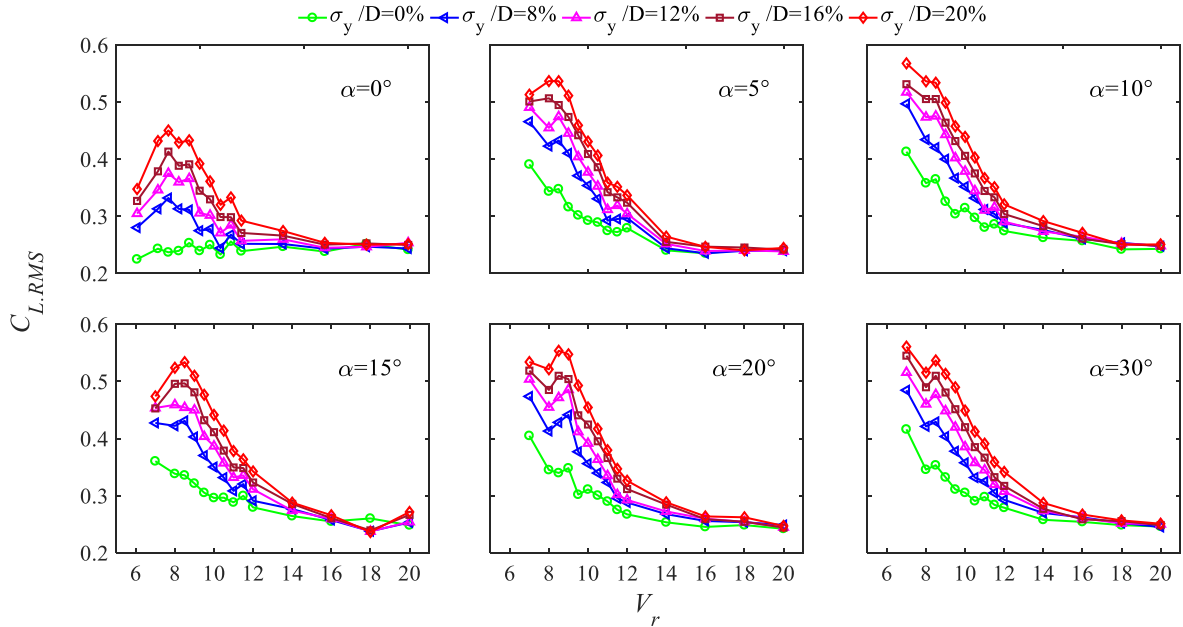

**Fig.9** Generalized RMS force coefficients.

*3.2.4 Altitudinal Lift*

The Altitudinal RMS lift coefficients, showing the altitudinal lift, are calculated by Equation (1) to instigate a better understanding of the force distribution**Error! Reference source not found.**. Fig.10 and Fig.11 present the altitudinal RMS lift in the VIV and Galloping regimes. In the VIV regime (Fig.10), the bi-polar change induced by transverse inclination is clear. The vertical case differs fundamentally from all others because the altitudinal lift is maximum at Level 7 (~$z/H$ = 0.8) and minimum at Level 1 (base). For inclined cases, the fundament behavioral change is the intensification of lift near the base, where it is maximum at Level 1 (base) and minimum at Level 10 (free-end). More importantly, the altitudinal lifts from Level 5 and above (mid-span) are identical for the vertical and incline cases. This means the force amplification observed in Fig.9 is exclusively attributed to the intensified force near the base, echoing with the phenomenological analysis from Fig.5 and Fig.6.

Fig.9 also reconfirms several other notions. First, a further increase of $α$ does not affect the unsteady forces, as the profiles of all inclined cases are nearly identical. Second, the Base Intensification does not see the effect of $σ_y/D$ because structure stiffness increasingly resists the free-vibration mode, so almost no motion is transferred to the base. Third, $α$ does not affect the free-end at all, not even an initial behavioral change.

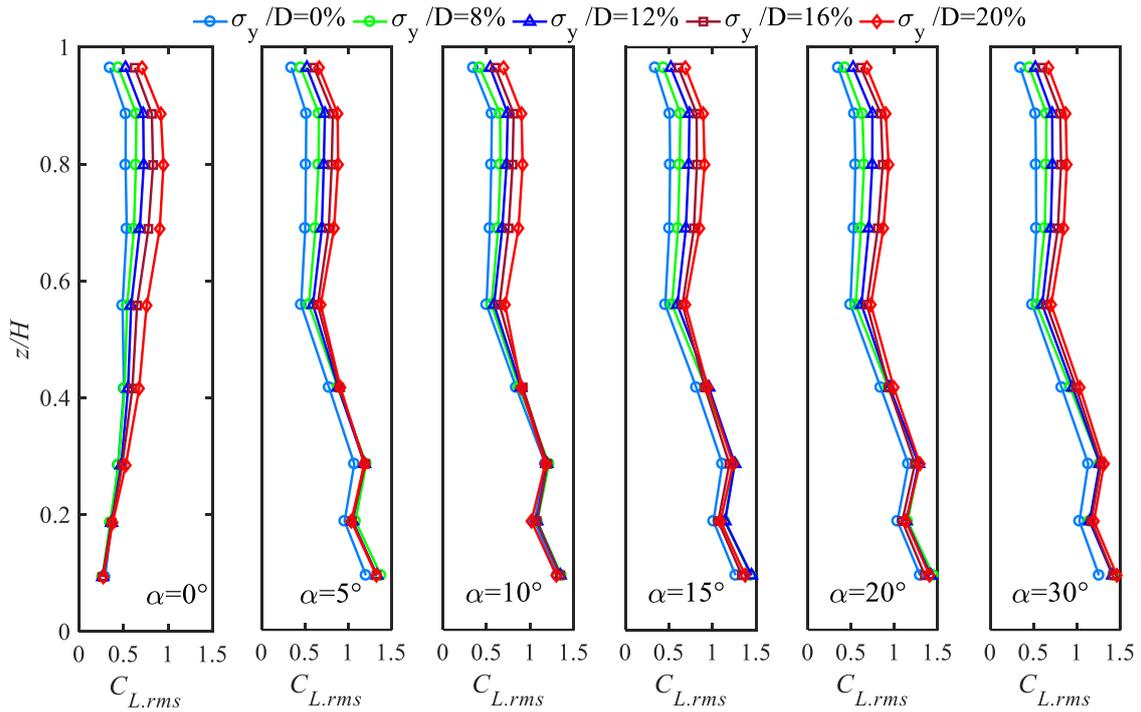

**Fig.10** Altitudinal RMS lift coefficient in the low wind speed ($V_r = 11$).

The observations translate to critical awareness in engineering practice. The Base Intensification's promotion of the VIV does not manifest as increased free-end motion, which is at least visibly detectable. Its effect is substantial at a structure's foundation because the unsteady force increases by as much as 4.3 times (from ~0.35 to 1.5 at Level 1), which may even be higher than the factor of safety in some designs. More alarmingly, this intensification is covert and hardly detectable in practice. One can easily imagine that Base Intensification induces a great cross-wind force that shears the lower half of a structure, which is typically more dangerous than an axial loading.

In the Galloping regime, Fig.11 shows that lift is uniform across the model height and unaffected by either α or $\sigma_y/D$. The collapse of all self-similar curves into a single curve reinforces the quasi-steady notion. Another interesting observation is made on the upper half: the maximum RMS lift for the upper half does not occur at the free-end but immediately below. This observation bears importance in the subsequent sections.

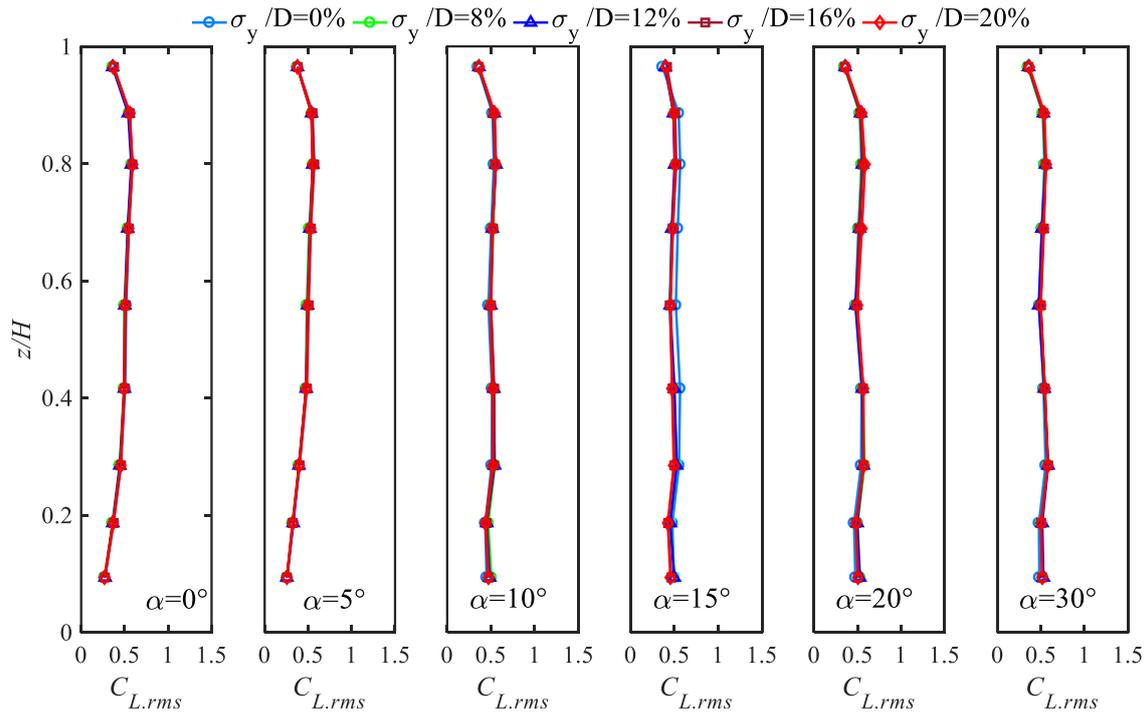

**Fig.11** Altitudinal RMS lift coefficient in the high wind speed ($V_r = 18$).

### 3.3 Power Spectral Density

*3.3.1  Generalized Force Spectra*

From force distribution, we now move on to spectral analysis. Fig.12 and Fig.13 present the generalized crosswind force spectra at α = 0°, 15°, and 30°. In the VIV regime (Fig.12), and for the static cases ($\sigma_y/D$=0%), only one peak corresponding to the Bérnard-Kármán vortex shedding, or more precisely, the locked-in shedding and structure frequencies, is observed (Bearman and Obasaju, 1982). Vibratory motion uniformly enhances the intensity of the primary peak, shifts it to a lower reduced frequency, or *Strouhal Number St = fD/U*, to ~0.09, and generates a second peak. Evidently, the twin peaks are merged, illustrating the complex and nonlinear entanglement between vortex shedding, natural vibration, and forced vibration. In the $\sigma_y/D=18\%$ case, even a high-frequency third peak can be observed. Finally, since the spectra mainly describe the vortex shedding, the Base Phenomenon, or $\alpha$, barely projects any influence.

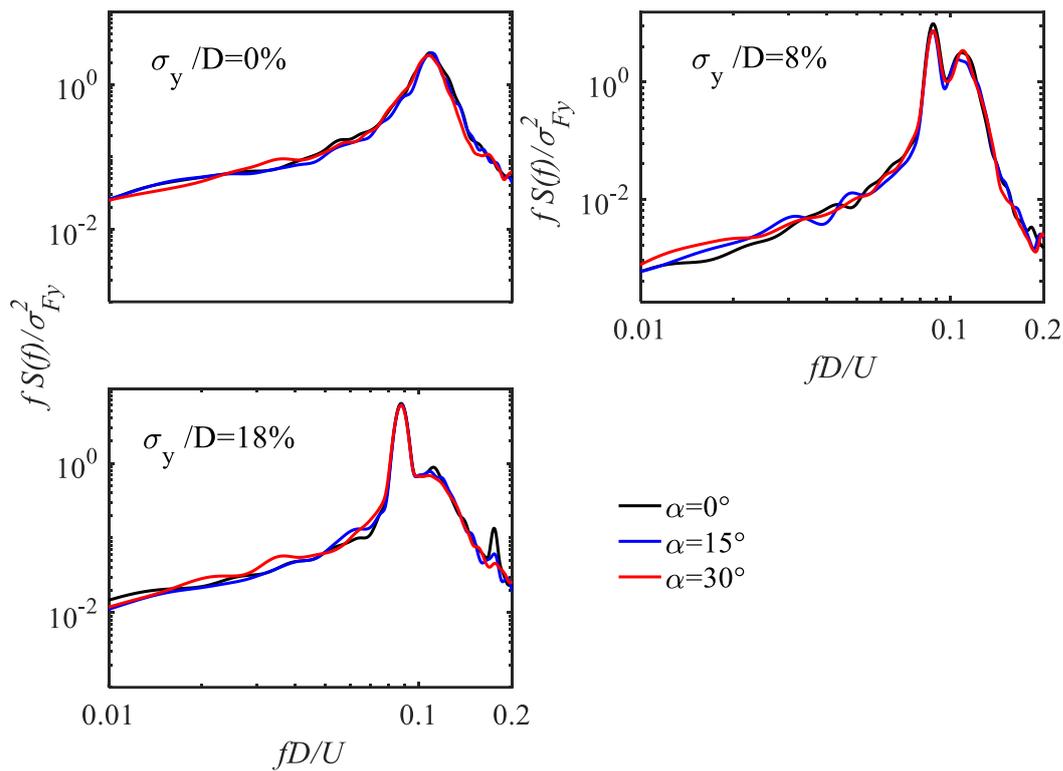

**Fig.12** Generalized force spectrum in the crosswind direction ($V_r = 11$).

One marked difference between the VIV and Galloping regimes (Fig.13) is the distinct separation of the energy concentrations. Here, the single peak in $\sigma_y/D=0\%$ describes only the Bérnard-Kármán vortex shedding. The forced vibration adds a low-frequency peak of less intensity, and no nonlinear entanglement is observed. Again, $\alpha$ barely projects any influence.

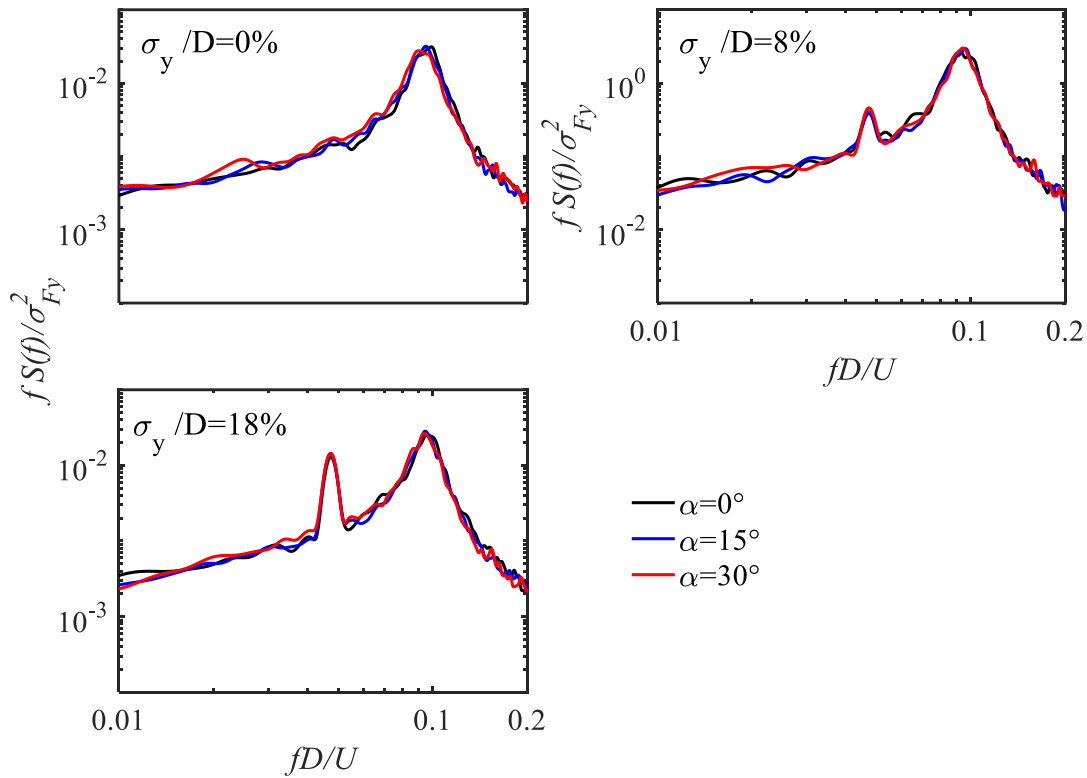

**Fig.13** Generalized force spectrum in the crosswind direction ($V_r = 18$).

*3.3.2 Pointwise Pressure Spectra*

Fig.14 and Fig.15 presents the pointwise pressure spectra at different altitudes. In the VIV regime (Fig.14), the merging of the primary $St\sim0.11$ peak becomes increasingly sharp from Levels 6 to 9, corresponding to the increasing dominance vortex shedding towards the free-end. Expectedly, Level 1 is minimally affected by the predominant vortex shedding.

Interestingly, a secondary peak at ~0.25 is visible at Levels 6 and 8 with $\alpha = 0°$, and both altitude and transverse inclination suppress this activity. The leading edge (A) is also uniformly more simplistic than the trailing edge (D). The downstream (C and D) exhibits a reduced primary peak and several secondary high-frequency peaks. This observation meets the conclusions from (Li et al., 2021b, 2021c), which spectrally characterized the prism wake. Separation, so shear layer dynamics, dominates the leading edge. But, towards the rear corners, besides the shear layers, reattachment and possibly re-separation also take place, not to mention the non-trivial presence of harmonic excitation in the prism base. Comparatively, the trailing edge has a more complex phenomenology as it involves several additional mechanisms. The spectra uniformly illustrate this energy redistribution into multiple peaks (Kareem, 1990; Kareem and Cermak, 1984).

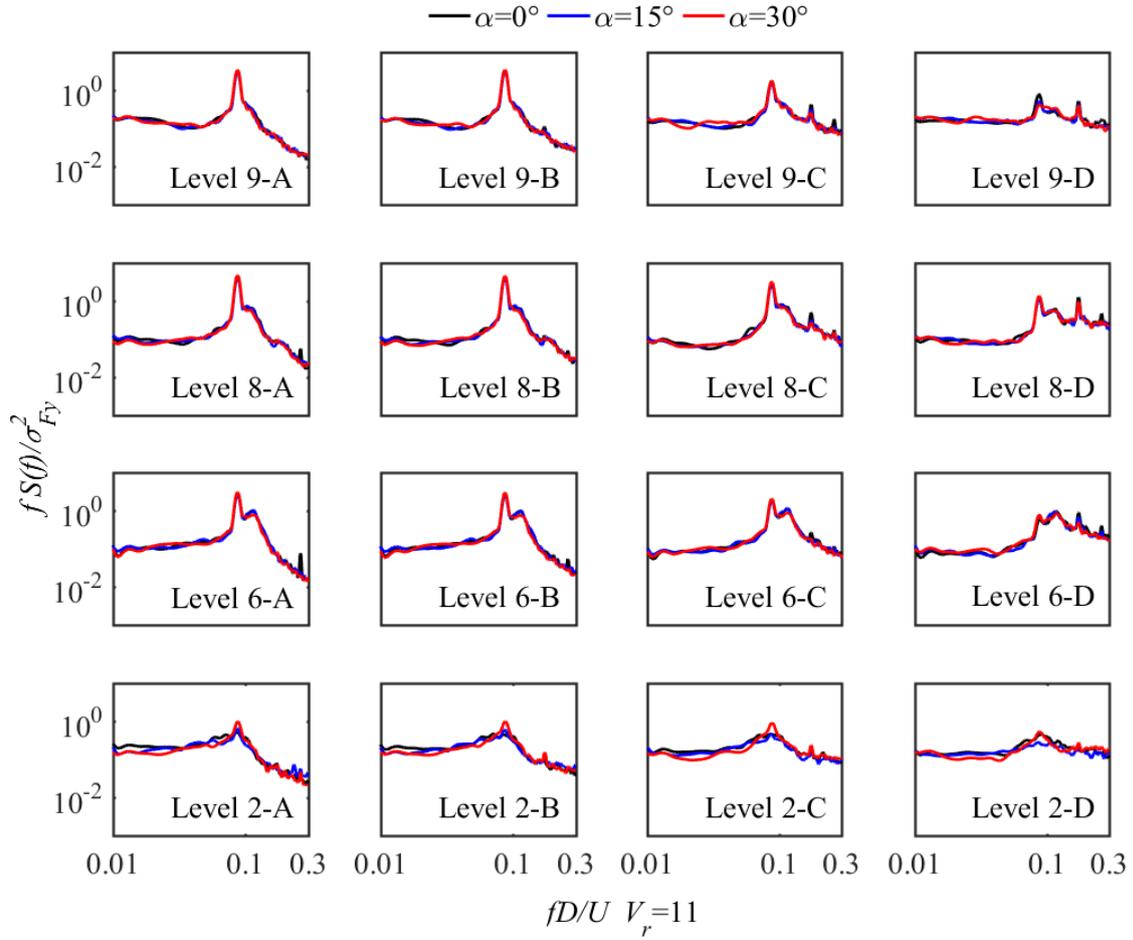

**Fig.14** Spectra of pointwise pressures on the side face of the test model at different levels($V_r = 11$).

On this note, transverse inclination's most prominent influence is at Level 2, as further accentuated by Fig.16. Take corner A as an example, increasing $\alpha$ from 15° to 30° sharpens the barely visible primary peak at $\alpha = 0°$, and shifts its frequency from ~0.08 to ~0.09. A secondary peak is also observed at ~0.2 for the inclined cases. Previously, we concluded that the Base Intensification is not related to shear layer dynamics, but it clearly promotes the Bérnard-Kármán vortex shedding near the base.

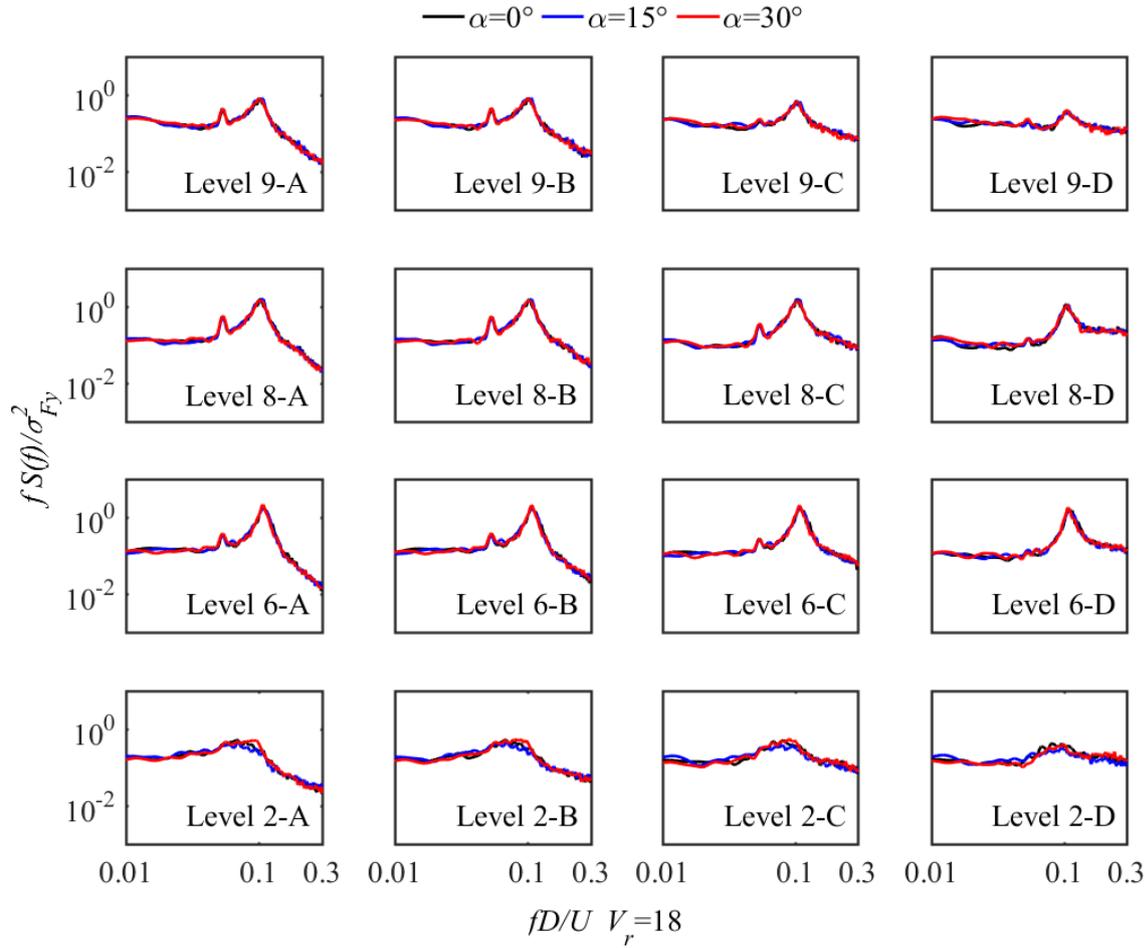

**Fig.15** Spectra of pointwise pressures on the side face of the test model at different levels ($V_r = 18$).

### 3.3.3 Suppression of Horseshoe Vortex

The purported paradox deserves an explanation. It is widely known that the fix-end horseshoe vortex disrupts shear layer activities, hence the formation of the Kármán vortex. This is why the spectral peak of $\alpha = 0°$ has a different frequency at St~0.08 and is much milder than higher levels. The Base Intensification impinges the horseshoe vortex, which effectively weakens the suppressor of the Kármán vortex. The suppression shifts the peak towards the shedding frequency ~0.11 (see Fig.17). So, Base Intensification promotes Bérnard-Kármán vortex shedding not by directly adding more kinetic energy, but by indirectly weakening its mitigatory opponent. Therefore, the force amplification observed in Section 3.2 is attributed to the unleashing of the Kármán vortex.

Finally, Fig.15 shows that transverse inclination does not affect the spectral content in the Galloping regime, agreeing with previous conclusions.

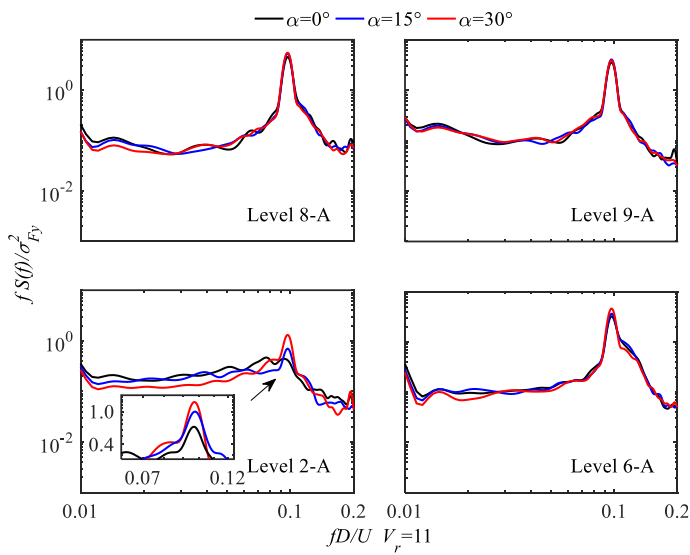 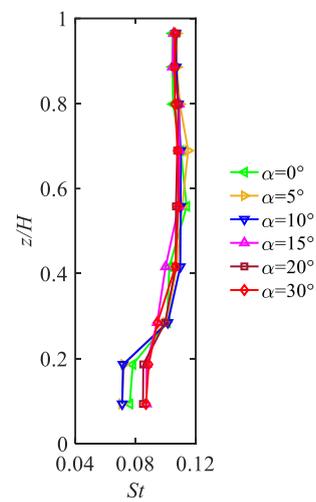

**Fig.16** Pointwise spectra analysis at points A at different heights.

**Fig.17** Altitudinal Strouhal numbers ($V_r = 18$).

On a different note, the effect of vibration amplitude is isolated and further investigated by parametrically measuring shedding frequency against wind speed for the $\sigma_y/D$=0% and 18% cases (see Fig.18). $St$ is linearly proportional to wind speed without an initial excitation. We observe no frequency lock-in, which means VIV does not take place. This observation agrees with the lift force in Fig.9. To this end, even though transverse inclination drastically changes the flow phenomenology and amplifies the crosswind loading on a structure base, the repercussions do not translate into visible VIV motion because the fix-end boundary condition reinforces the stiffness. Nevertheless, this can be a double-edged sword as stress builds up without elastic mitigation, so failures, if they are to occur, will come more cataclysmic, irreversible, and without any warning signs.

By contrast, the shedding frequency locks with the natural frequency of the structure (7.8 *Hz*) with $\sigma_y/D$=18%. The observation accentuates the deciding role of initial perturbation in the activation of VIV. Transverse inclination also does not interfere with the VIV motion, which typically concerns the free-end instead of the fix-end. After the lock-in, the linear proportionality is suddenly restored for the Galloping regime, attesting to the suitability of the quasi-steady theory therein. The vivid manifestation of the lock-in also agrees with several previous studies(Bearman and Obasaju, 1982; Chen et al., 2021b, 2017).

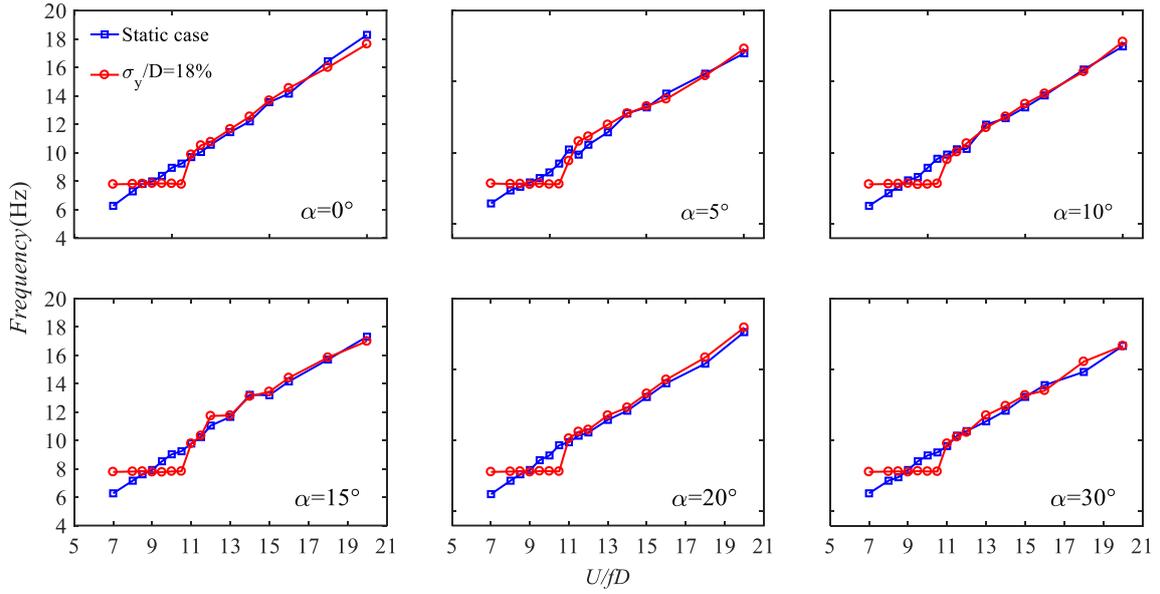

**Fig.18** Comparison of static and forced-vibration tests with vortex shedding frequency.

## 4. Aerodynamic Damping

### 4.1 Identification of Aerodynamic Damping

Aerodynamic damping plays a key role in almost all regimes of crosswind vibration, including VIV, Galloping, Flutter, etc. (Abdel-Rohman, n.d.; Dai et al., 2017, 2014; Steckley et al., 1990; Vickery and Steckley, 1993). This section is dedicated to the aerodynamic damping of transversely inclined structures.

The equation of motion of a slender prism under winding loading per unit height at an elevation $z$ above the ground is expressed as

$$m(z)\{\ddot{y}(z,t) + 2\xi_y \omega_y \dot{y}(z,t) + \omega_y^2 y(z,t)\} = P(z,t) \qquad (6)$$

where $m(z)$ is the mass at height $z$, $\xi_y$ is the damping of the structure, $\omega_y$ is the circular frequency of forced vibration, $y(z,t), \dot{y}(z,t), \ddot{y}(z,t)$ are the displacement, speed and acceleration of the prism, respectively; $P(z,t)$ is the wind-induced force at the height of $z$, which can be expressed as:

$$P(z,t) = 0.5\rho U^2 [C_M(z,t) + C_L(z,t)]D \tag{7}$$

where $C_L(z,t)$ and $C_M(z,t)$ are the local crosswind force and the motion-induced force coefficients at height $z$, respectively. $\rho$ is the density of air; $U$ is the wind speed at point of reference.

In the forced-vibration test, the prism is driven harmonically and the time history of tip response is $y(t) = \hat{y}\cos 2\pi f t$, where $\hat{y}$ is the maximum standard derivation response. The unsteady aerodynamic force can be decomposed into aerodynamic damping component and aerodynamic stiffness component, which can be expressed as:

$$P(z,t) = Q_k \frac{y(t)}{\hat{y}} + Q_d \frac{\dot{y}(t)}{2\pi f \hat{y}} = Q_k \cos 2\pi f t - Q_d \sin 2\pi f t \tag{8}$$

where the aerodynamic stiffness force coefficient $Q_k$ and aerodynamic damping force coefficient $Q_d$ are expressed as:

$$Q_k = \frac{2}{T}\int_0^T P(z,t)\frac{y(t)}{\hat{y}}dt \tag{9}$$

$$Q_d = \frac{2}{T}\int_0^T P(z,t)\frac{\dot{y}(t)}{2\pi f \hat{y}}dt \tag{10}$$

One shall note that $Q_k$ and $Q_d$ are aerodynamic stiffness component and aerodynamic damping component, respectively, which are in-phase with response and velocity. $Q_k$ is the real part of the motion-induced force, and $Q_d$ is the imaginary part.

By substituting the aforenoted equations, the local aerodynamic damping force coefficient $\chi(z)$ is expressed as

$$\chi(z) = \frac{2Q_d}{\rho D U^2} = \frac{4\int_0^T P(z,t)\frac{\dot{y}(t)}{2\pi f \hat{y}}dt}{\rho D U^2 T} \tag{11}$$

and the local aerodynamic stiffness force coefficient $\beta(z)$ as

$$\beta(z) = \frac{2Q_k}{\rho D U^2} = \frac{4\int_0^T P(z,t)\frac{y(t)}{\hat{y}}dt}{\rho D U^2 T} \tag{12}$$

The generalized aerodynamic damping force coefficient $\kappa$ and generalized aerodynamic stiffness force coefficient β can be determined by integrating $\chi(z)$ and $\beta(z)$ along the model height and expressed by

$$\kappa = \frac{\int_0^H \chi(z)\phi(z)dz}{H} \tag{13}$$

$$\beta = \frac{\int_0^H \beta(z)\phi(z)dz}{H} \tag{14}$$

where $\phi(z)$ is the linear mode shape in a forced vibration test ($\phi(z) = z/H$). It should be noted that the normalized damping coefficient $\chi$ is

$$\chi = \xi_b/\eta = -\frac{3}{4}\frac{D}{\hat{y}}\left(\frac{U}{Dw_v}\right)^2 \kappa \tag{15}$$

where $\eta = \rho D^2/m_v$, $\xi_b$ is aerodynamic damping ratio, $m_v$ is the mass per unit height. Since, the aerodynamic damping ratio $\xi_b$ can be expressed as

$$\xi_b = -\frac{3}{4}\frac{D}{\hat{y}}\left(\frac{\rho D^2}{m_v}\right)\left(\frac{U}{Dw_v}\right)^2 \kappa \tag{16}$$

**4.2 Normalized Damping**

Following the mathematical rendering, Fig.19 presents the normalized aerodynamic damping $\chi$ obtained from Eqs. (13) and (15) at different $\sigma_y/D$, $V_r$, and $\alpha$. In general, the normalized $\chi$ is positive in the VIV regime, manifesting its self-containing nature. $\chi$ is negative in the Galloping regime, displaying its diverging tendency. So, the zero-ordinate marks the end of the lock-in and the beginning of unconstrained vibration afterward. In terms of inter-dependence, $\alpha$ only peripherally alters $\chi$ as its distributions are self-similar. Likewise, $\sigma_y/D$ only slightly increases $\chi$ in the lock-in range, promoting the aerodynamic dissipation against VIV and extending its range to a higher $V_r$. Neither $\alpha$ nor $\sigma_y/D$ affects the Galloping regime significantly. Finally, we take note of the plummet of $\chi$ at $\alpha = 30°$ and $\sigma_y/D = 10\%$ in both figures. This singularity is deemed trivial and does not affect the overall trend.

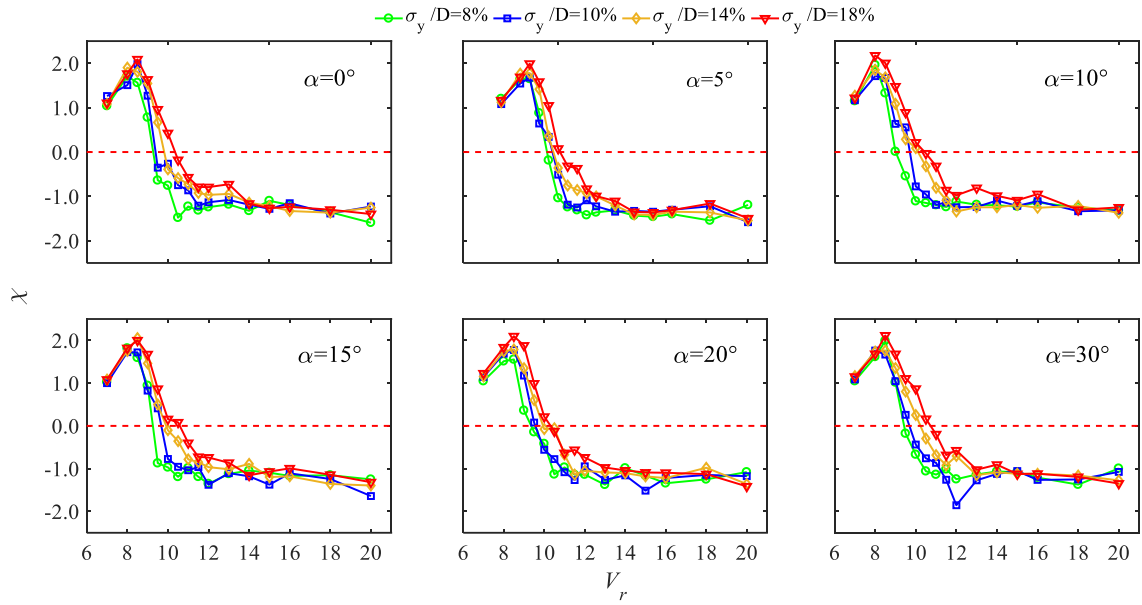

**Fig.19** Normalized aerodynamic damping for transversely inclined prisms.

### 4.3 Altitudinal Damping

According to Eq. (11), we calculated the altitudinal aerodynamic damping for an intermediate $\sigma_y/D = 14\%$ to better dissect normalized coefficient. Fig.20 shows the normalized $\chi$ is dominated by the altitudinal $\chi(z)$ of the free-end. At Levels 6-9, $\chi(z)$ generally resembles $\chi$ from Fig.19 except the showing a clear convex inflexion (minima) immediately after lock-in. The observation outlines the most aerodynamically unfavorable state, the merge of VIV and Galloping, or VIV-Galloping interaction. This audaciously unconstrained state has been observed and studied by(Williamson and Roshko, 1988) and (Mannini et al., 2015a).

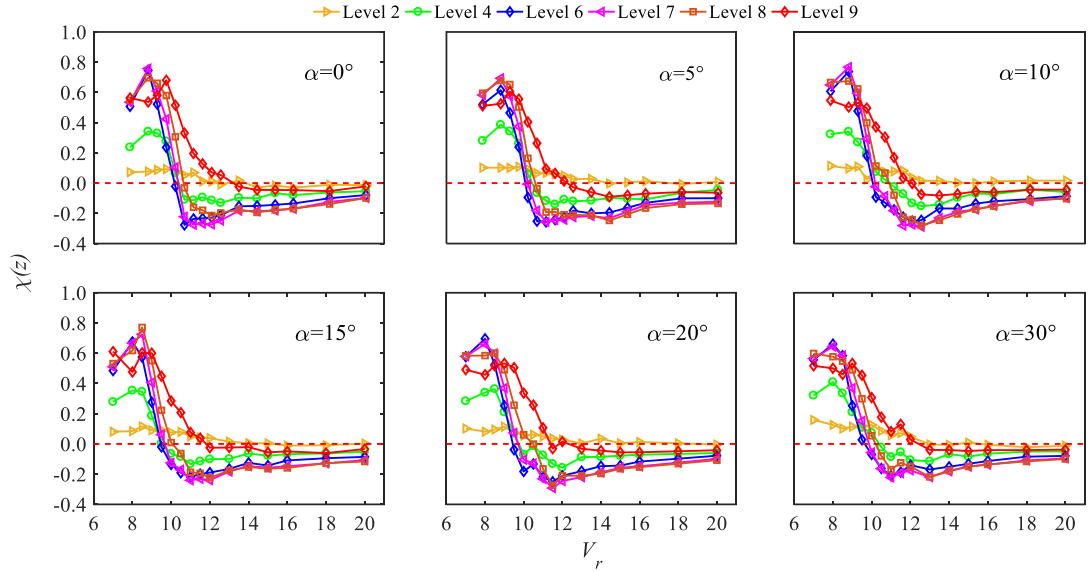

**Fig.20** Altitudinal aerodynamic damping for transversely inclined structures.

At Level 2, $\chi(z)$ hovers around zero and never really becomes negative in the Galloping regime. So, galloping will never occur at the prism base. Another unexpected observation is that the maximum $\chi(z)$ does not occur at Level 9, but rather at Levels 7 and 8. Readers are also remined by similar observations made on the altitudinal lift (see Fig.10 and Fig.11). The explanation is as follows.

The oddities of Level 2 and Level 9 are attributed to different aerodynamic mechanisms. As is well-known, the predominant driving mechanism is the Bérnard-Kármán vortex shedding. The first exception is at the model base, where the Kármán vortex is overshadowed by the horseshoe vortex. As the result, the entire array of aerodynamic parameters, including the aforementioned mean pressure, lift, Strouhal number, and aerodynamic damping are fundamentally different from higher levels. As for the free-end, the Kármán vortex is affected by three-dimensional effects like the axial flow(Hu et al., 2016). However, the degree is not as overwhelming as fixed end. Consequently, the aerodynamic parameters maintain dominance with a compromised intensity.

Perhaps this is also a good point to elucidate the aerodynamic effect of the transverse inclination. As all evidence suggest, the transverse inclination induces a bi-polar step-change to the aerodynamic behaviors of the prism. A prescription of any degrees of *α* projects a notable impact to the bottom half of the model, while the upper half remains unaffected. The impact consists of drastically enhanced base RMS lift, elevated base Strouhal number, and here the extended range of positive $\chi(z)$. All of these suggest a universal tenet: the transverse inclination intensifies fix-end aerodynamic phenomena like the horseshoe vortex, causing a more erratic wind field near, and stronger loadings onto, the structure base. Yet, once the initial perturbation, the degree of the intensification is influenced by neither the angle of inclination nor the amplitude of tip vibration. The practical implications are that structural inclinations of any orientation, as it easily changes with the wind attack angle, may cause unpredictable wind environment at the pedestrian level, and demand more safety reinforcement near the build base. Therefore, in addition to the existing observations in the literature, it is remarked that the *Base Intensification* phenomenon will add another layer of complexity to the aerodynamics of inclined structures.

## 5. Conclusion

This paper investigates the unsteady aerodynamic characteristic of transversely inclined square prisms using the forced-vibration technique, in which different wind speeds, inclination angles, and oscillation amplitudes have been tri-parametrically tested. Results show that transverse inclination induces a bi-polar, once-for-all change in the flow morphology and aerodynamic characteristics. A transverse inclination of any degree, whether by design or change of attack angle, impinges the fix-end horseshoe vortex and breaks the wake symmetry. At a structure's base, the direct repercussion is the unleashing of the Bérnard-Kármán vortex shedding, which increases unsteady lift by as much as 4.3 times. Without the horseshoe vortex, the structure's free-end only marginally feels the effect. To this end, the phenomenological change is termed as the *Base Intensification*. Upon the initial activation, the inclined structure becomes independent of inclination angles and oscillation amplitudes and depends solely on wind speeds. Specifically, the aerodynamic characteristics are altered in the low-speed VIV regime and intact in the quasi-steady Galloping regime.

Finally, some engineering implications deserve a discussion. Due to the *Base Intensification* phenomenon, the safety of inclined structures, regardless of incline orientation, demands more engineering caution. Intensified pedestrian-level winds may result from the unleashed vortex shedding. Moreover, the excessive aerodynamic load adds massive stress to a structure's base because the stiffness prevents any elastic dissipation. Since the intensification does not translate to any motion, not only will detection become trickier, but if failures are also to occur, they will be more drastic and without any warnings. The 4.3-time force amplification is greater than the safety factor of many standard designs, so the issue deserved immediate attention.

**Acknowledgement**

The authors appreciate the testing facility, as well as the technical assistance provided by the CLP Power Wind/Wave Tunnel Facility at the Hong Kong University of Science and Technology. The authors also thank the Design and Manufacturing Services Facility (Electrical and Mechanical Fabrication Unit) of the Hong Kong University of Science and Technology for their help in manufacturing the test rig of the forced vibration wind tunnel test system.

# References

Abdel-Rohman, M., n.d. Effect of Unsteady Wind Flow on Galloping of Tall Prismatic Structures 22.

Bearman, P.W., Obasaju, E.D., 1982. An experimental study of pressure fluctuations on fixed and oscillating square-section cylinders. J. Fluid Mech. 119, 297–321. https://doi.org/10.1017/S0022112082001360

Blevins, R.D., Saunders, H., 1979. Flow Induced Vibration. J. Mech. Des. 101, 6–6. https://doi.org/10.1115/1.3454027

Carassale, L., Freda, A., Marrè-Brunenghi, M., 2014. Experimental investigation on the aerodynamic behavior of square cylinders with rounded corners. J. Fluids Struct. 44, 195–204. https://doi.org/10.1016/j.jfluidstructs.2013.10.010

Chen, Z., Fu, X., Xu, Y., Li, C.Y., Kim, B., Tse, K.T., 2021a. A perspective on the aerodynamics and aeroelasticity of tapering: Partial reattachment. J. Wind Eng. Ind. Aerodyn. 212, 104590. https://doi.org/10.1016/j.jweia.2021.104590

Chen, Z., Huang, H., Tse, K.T., Xu, Y., Li, C.Y., 2020. Characteristics of unsteady aerodynamic forces on an aeroelastic prism: A comparative study. J. Wind Eng. Ind. Aerodyn. 205, 104325. https://doi.org/10.1016/j.jweia.2020.104325

Chen, Z., Huang, H., Xu, Y., Tse, K.T., Kim, B., Wang, Y., 2021b. Unsteady aerodynamics on a tapered prism under forced excitation. Eng. Struct. 240, 112387. https://doi.org/10.1016/j.engstruct.2021.112387

Chen, Z., Tse, K.T., Hu, G., Kwok, K.C.S., 2018a. Experimental and theoretical investigation of galloping of transversely inclined slender prisms. Nonlinear Dyn. 91, 1023–1040. https://doi.org/10.1007/s11071-017-3926-y

Chen, Z., Tse, K.T., Kwok, K.C.S., 2017. Unsteady pressure measurements on an oscillating slender prism using a forced vibration technique. J. Wind Eng. Ind. Aerodyn. 170, 81–93. https://doi.org/10.1016/j.jweia.2017.08.004

Chen, Z., Tse, K.T., Kwok, K.C.S., Kareem, A., 2018b. Aerodynamic damping of inclined slender prisms. J. Wind Eng. Ind. Aerodyn. 177, 79–91. https://doi.org/10.1016/j.jweia.2018.04.016

Chen, Z., Tse, K.T., Kwok, K.C.S., Kareem, A., Kim, B., 2021c. Measurement of unsteady aerodynamic force on a galloping prism in a turbulent flow: A hybrid aeroelastic-pressure balance. J. Fluids Struct. 102, 103232. https://doi.org/10.1016/j.jfluidstructs.2021.103232

Cooper, K.R., Nakayama, M., Sasaki, Y., Fediw, A.A., Resende-Ide, S., Zan, S.J., 1997. Unsteady aerodynamic force measurements on a super-tall building with a tapered cross section. J. Wind Eng. Ind. Aerodyn. 72, 199–212. https://doi.org/10.1016/S0167-6105(97)00258-4

Dai, H.L., Abdelkefi, A., Wang, L., Liu, W.B., 2014. Control of cross-flow-induced vibrations of square cylinders using linear and nonlinear delayed feedbacks. Nonlinear Dyn. 78, 907–919. https://doi.org/10.1007/s11071-014-1485-z

Dai, H.L., Abdelmoula, H., Abdelkefi, A., Wang, L., 2017. Towards control of cross-flow-induced vibrations based on energy harvesting. Nonlinear Dyn. 88, 2329–2346. https://doi.org/10.1007/s11071-017-3380-x

Flemming, F., Williamson, C.H.K., 2005. Vortex-induced vibrations of a pivoted cylinder. J. Fluid Mech. 522, 215–252. https://doi.org/10.1017/S0022112004001831

He, Y., Zhang, L., Chen, Z., Li, C.Y., 2022. A framework of structural damage detection for civil structures using a combined multi-scale convolutional neural network and echo state network. Eng. Comput. https://doi.org/10.1007/s00366-021-01584-4

Holmes, J.D., 2018. Wind Loading of Structures, 0 ed. CRC Press. https://doi.org/10.1201/b18029

Hu, G., Tse, K.T., Chen, Z.S., Kwok, K.C.S., 2017. Particle Image Velocimetry measurement of flow around an inclined square cylinder. J. Wind Eng. Ind. Aerodyn. 168, 134–140. https://doi.org/10.1016/j.jweia.2017.06.001

Hu, G., Tse, K.T., Kwok, K.C.S., 2016. Aerodynamic mechanisms of galloping of an inclined square cylinder. J. Wind Eng. Ind. Aerodyn. 148, 6–17. https://doi.org/10.1016/j.jweia.2015.10.011

Hu, G., Tse, K.T., Kwok, K.C.S., 2015. Galloping of forward and backward inclined slender square cylinders. J. Wind Eng. Ind. Aerodyn. 142, 232–245. https://doi.org/10.1016/j.jweia.2015.04.010

Hu, Gang, Tse, K.T., Kwok, K.C.S., Chen, Z.S., 2015. Pressure measurements on inclined square prisms. Wind Struct. 21, 383–405. https://doi.org/10.12989/WAS.2015.21.4.383

Hui, Y., Yuan, K., Chen, Z., Yang, Q., 2019. Characteristics of aerodynamic forces on high-rise buildings with various façade appurtenances. J. Wind Eng. Ind. Aerodyn. 191, 76–90. https://doi.org/10.1016/j.jweia.2019.06.002

Kareem, A., 1990. Measurements of pressure and force fields on building models in simulated atmospheric flows. Sixth US Natl. Conf. Wind Eng. 36, 589–599. https://doi.org/10.1016/0167-6105(90)90341-9

Kareem, A., Cermak, J.E., 1984. Pressure fluctuations on a square building model in boundary-layer flows. J. Wind Eng. Ind. Aerodyn. 16, 17–41. https://doi.org/10.1016/0167-6105(84)90047-3

Kim, Y.C., Lo, Y.L., Chang, C.H., 2018. Characteristics of unsteady pressures on slender tall building. J. Wind Eng. Ind. Aerodyn. 174, 344–357. https://doi.org/10.1016/j.jweia.2018.01.027

Li, C.Y., Chen, Z., Tse, T.K.T., Weerasuriya, A.U., Zhang, X., Fu, Y., Lin, X., 2022. A parametric and feasibility study for data sampling of the dynamic mode decomposition: range, resolution, and universal convergence states. Nonlinear Dyn. https://doi.org/10.1007/s11071-021-07167-8

Li, C.Y., Chen, Z., Tse, T.K.T., Weerasuriya, A.U., Zhang, X., Fu, Y., Lin, X., 2021a. Establishing direct phenomenological connections between fluid and structure by the Koopman-Linearly Time-Invariant analysis. Phys. Fluids 33, 121707. https://doi.org/10.1063/5.0075664

Li, C.Y., Chen, Z., Tse, T.K.T., Weerasuriya, A.U., Zhang, X., Fu, Y., Lin, X., 2021b. Spectral Characterization by the Koopman Linearly-Time-Invariant Analysis: Constitutive Fluid-Structure Correspondence. ArXiv211202985 Phys.

Li, C.Y., Chen, Z., Tse, T.K.T., Weerasuriya, A.U., Zhang, X., Fu, Y., Lin, X., 2021c. Spectral Characterization by the Koopman Linearly-Time-Invariant Analysis: Phenomenological Fluid-Structure Correspondence and Its Origins. ArXiv211203029 Phys.

Li, C.Y., Tse, T.K.T., Hu, G., 2020. Dynamic Mode Decomposition on pressure flow field analysis: Flow field reconstruction, accuracy, and practical significance. J. Wind Eng.

Ind. Aerodyn. 205, 104278. https://doi.org/10.1016/j.jweia.2020.104278

Lin, N., Letchford, C., Tamura, Y., Liang, B., Nakamura, O., 2005. Characteristics of wind forces acting on tall buildings. J. Wind Eng. Ind. Aerodyn. 93, 217–242. https://doi.org/10.1016/j.jweia.2004.12.001

Mannini, C., Marra, A.M., Bartoli, G., 2015a. Experimental investigation on VIV-galloping interaction of a rectangular 3:2 cylinder. Meccanica 50, 841–853. https://doi.org/10.1007/s11012-014-0025-8

Mannini, C., Marra, A.M., Bartoli, G., 2015b. Experimental investigation on VIV-galloping interaction of a rectangular 3:2 cylinder. Meccanica 50, 841–853. https://doi.org/10.1007/s11012-014-0025-8

Mannini, C., Massai, T., Maria Marra, A., Bartoli, G., 2017. Interference of Vortex-Induced Vibration and Galloping: Experiments and Mathematical Modelling. Procedia Eng. 199, 3133–3138. https://doi.org/10.1016/j.proeng.2017.09.566

Païdoussis, M.P., Price, S.J., Langre, E. de, 2013. Fluid-structure interactions: cross-flow-induced instabilities. Cambridge university press, Cambridge.

Parkinson, G., 1989. Phenomena and modelling of flow-induced vibrations of bluff bodies. Prog. Aerosp. Sci. 26, 169–224. https://doi.org/10.1016/0376-0421(89)90008-0

Raissi, M., Wang, Z., Triantafyllou, M.S., Karniadakis, G.E., 2019. Deep learning of vortex-induced vibrations. J. Fluid Mech. 861, 119–137. https://doi.org/10.1017/jfm.2018.872

Steckley, A., Vickery, B.J., Isyumov, N., 1990. On the measurement of motion induced forces on models in turbulent shear flow. J. Wind Eng. Ind. Aerodyn. 36, 339–350. https://doi.org/10.1016/0167-6105(90)90318-7

Tanaka, H., Tamura, Y., Ohtake, K., Nakai, M., Chul Kim, Y., 2012. Experimental investigation of aerodynamic forces and wind pressures acting on tall buildings with various unconventional configurations. J. Wind Eng. Ind. Aerodyn. 107–108, 179–191. https://doi.org/10.1016/j.jweia.2012.04.014

Unal, M.F., Rockwell, D., 1988a. On vortex formation from a cylinder. Part 1. The initial instability. J. Fluid Mech. 190, 491–512. https://doi.org/10.1017/S0022112088001429

Unal, M.F., Rockwell, D., 1988b. On vortex formation from a cylinder. Part 2. Control by

splitter-plate interference. J. Fluid Mech. 190, 513–529. https://doi.org/10.1017/S0022112088001430

Vickery, B.J., Steckley, A., 1993. Aerodynamic damping and vortex excitation on an oscillating prism in turbulent shear flow. J. Wind Eng. Ind. Aerodyn. 49, 121–140. https://doi.org/10.1016/0167-6105(93)90009-D

Williamson, C.H.K., Roshko, A., 1988. Vortex formation in the wake of an oscillating cylinder. J. Fluids Struct. 2, 355–381. https://doi.org/10.1016/S0889-9746(88)90058-8

Zhang, X., Weerasuriya, A.U., Wang, J., Li, C.Y., Chen, Z., Tse, K.T., Hang, J., 2022. Cross-ventilation of a generic building with various configurations of external and internal openings. Build. Environ. 207, 108447. https://doi.org/10.1016/j.buildenv.2021.108447

Zhang, Xinyue, Weerasuriya, A.U., Zhang, Xuelin, Tse, K.T., Lu, B., Li, C.Y., Liu, C.-H., 2020. Pedestrian wind comfort near a super-tall building with various configurations in an urban-like setting. Build. Simul. 13, 1385–1408. https://doi.org/10.1007/s12273-020-0658-6

Zou, L., Li, F., Song, J., Shi, T., Liang, S., Mercan, O., 2020. Investigation of torsional aeroelastic effects on high-rise buildings using forced vibration wind tunnel tests. J. Wind Eng. Ind. Aerodyn. 200, 104158. https://doi.org/10.1016/j.jweia.2020.104158

**Funding**

The work was supported by the National Natural Science Foundation of China (Grant No.: 51908090), the Fundamental Research Funds for the Central Universities (Project No.: 2019CDXYTM0032), the Natural Science Foundation of Chongqing, China (Grant No.: cstc2019jcyj-msxm0639, cstc2020jcyj-msxmX0921), the Key project of Technological Innovation and Application Development in Chongqing (Grant No.: cstc2019jscx-gksb0188).

**Conflict of Interest**

The authors declare that they have no conflict of interest.

**Availability of Data and Material**

The datasets generated during and/or analyzed during the current work are restricted by provisions of the funding source but are available from the corresponding author on reasonable request.

**Code Availability**

The custom code used during and/or analyzed during the current work are restricted by provisions of the funding source.

**Author Contributions**

All authors contributed to the study's conception and design. Funding, project management, and supervision were performed by Jianmin Hua and Zengshun Chen. Material preparation, data collection, and formal analysis were led by Cruz Y. Li, Zengshun Chen, and assisted by Jie Bai, Yemeng Xu and Xuanyi Xue. The first draft of the manuscript was written by Cruz Y. Li, and all authors commented on previous versions of the manuscript. All authors read, contributed, and approved the final manuscript.

**Compliance with Ethical Standards**

All procedures performed in this work were in accordance with the ethical standards of the institutional and/or national research committee and with the 1964 Helsinki declaration and its later amendments or comparable ethical standards.

**Consent to Participate**

Informed consent was obtained from all individual participants included in the study.

**Consent for Publication**

Publication consent was obtained from all individual participants included in the study.